\documentclass[reqno,11pt]{amsart}

\usepackage{epsf,amssymb}

\def\be{\begin{equation}}
\def\ee{\end{equation}}
\def\ba{\begin{align}}
\def\bm{\begin{multline}}
\def\bfig{\begin{figure}[htb]}
\def\efig{\end{figure}}
\newcommand{\paper}[1]{{\it #1}, }
\newcommand{\journal}[4]{#1 #2, #3 (#4)}
\newcommand{\CMP}{Commun.\ Math.\ Phys.}
\newcommand{\HPA}{Helv.\ Phys.\ Acta}
\newcommand{\JSP}{J.\ Stat.\ Phys.}

\newtheorem{theorem}{Theorem}%[section]

\newtheorem{lemma}[theorem]{Lemma}

\newcommand{\nn}{\nonumber}

\renewcommand{\leq}{\;\leqslant\;}
\renewcommand{\geq}{\;\geqslant\;}

\newcommand{\dd}{{\rm d}}
\newcommand{\e}[1]{\,{\rm e}^{#1}\,}
\newcommand{\ii}{{\rm i}}
\newcommand{\sumtwo}[2]{\sum_{\substack{#1 \\ #2}}}
\newcommand{\sumthree}[3]{\sum_{\substack{#1 \\ #2 \\ #3}}}

\newcommand{\mintwo}[2]{\min_{\substack{#1 \\ #2}}}

\newcommand{\lowerint}[1]{\lfloor #1 \rfloor}

\def\Tr{{\operatorname{Tr\,}}}

\def\bra #1{\langle#1 |\,}
\def\ket #1{\,|#1 \rangle}

\newcommand{\const}{{\text{\rm const}}}
\newcommand{\upchi}{\raise 2pt \hbox{$\chi$}}
\newcommand{\Nc}{N_{\rm c}}
\newcommand{\Ne}{N_{\rm e}}
\newcommand{\muc}{\mu_{\rm c}}
\newcommand{\mue}{\mu_{\rm e}}
\newcommand{\rhoc}{\rho_{\rm c}}
\newcommand{\rhoe}{\rho_{\rm e}}
\def\writefig#1 #2 #3 {\rlap{\kern #1 truecm \raise #2 truecm
\hbox{#3}}}
\def\figtext#1{\smash{\hbox{#1}} \vspace{-5mm}}
\newcommand{\caD}{{\mathcal D}}

\newcommand{\caF}{{\mathcal F}}

\newcommand{\caH}{{\mathcal H}}

\newcommand{\bbN}{{\mathbb N}}

\newcommand{\bbR}{{\mathbb R}}

\newcommand{\bbZ}{{\mathbb Z}}

\begin{document}

\title{The Falicov-Kimball model}

\author[Christian Gruber, Daniel Ueltschi]{Christian Gruber and Daniel Ueltschi}

\address{Christian Gruber \hfill\newline
Institut de Physique Th\'eorique \hfill\newline
\'Ecole Polytechnique F\'ed\'erale \hfill\newline
CH--1015 Lausanne, Switzerland}
\email{christian.gruber@epfl.ch}

\address{Daniel Ueltschi \hfill\newline
Department of Mathematics \hfill\newline
University of Arizona \hfill\newline
Tucson, AZ 85721, USA \newline\indent
{\rm http://math.arizona.edu/$\sim$ueltschi}}
\email{ueltschi@math.arizona.edu}


\thanks{{\it 2000 Math.\ Subj.\ Class.:} 82B10, 82B20, 82B26.}


\maketitle

\section{A brief history}

The ``Falicov-Kimball model'' was first considered by Hubbard and Gutzwiller in 1963--65
as a simplification of the Hubbard model. Falicov and Kimball introduced in 1969 a model
that included a few extra complications, in order to investigate metal-insulator phase
transitions in rare-earth materials and transition-metal compounds. Experimental data
suggested that this transition is due to the interactions between electrons in two
electronic states: non-localized states (itinerant electrons), and states that are
localized around the sites corresponding to the metallic ions of the crystal (static
electrons).

A tight-binding approximation leads to a model defined on a lattice (the crystal) and two species of particles are considered.
The first species consists of spinless quantum fermions (we refer to them as
``electrons''), and the second species consists of localized holes or electrons
(``classical particles''). Electrons hop between nearest-neighbor sites but classical
particles do not. Both species obey Fermi statistics (in particular, the Pauli exclusion principle
prevents more than one particle of a given species to occupy the same site).
Interactions are on-site and thus involve particles of different species; they can be
repulsive or attractive.

The very simplicity of the model allows for a broad range of applications. It was
studied in the context of mixed valence systems, binary alloys, and crystal formation.
Adding a magnetic field yields the flux phase problem. The Falicov-Kimball model can also
be viewed as the simplest model where quantum particles interact with classical fields.

The fifteen years following the introduction of the model saw studies based on approximate
methods, such as Green's function techniques, that gave rise to a lot of confusion. A
breakthrough occurred in 1986 when Brandt and Schmidt, and Kennedy and Lieb, proposed the
first rigorous results. In particular, Kennedy and Lieb showed in their beautiful paper that
the electrons create an effective interaction between the classical particles, and that a phase transition takes place for
any value of the coupling constant, provided the temperature is low enough.

Many studies by mathematical-physicists followed and several results are presented in this
short survey. Recent years have seen an increase of interest by condensed matter
physicists. Only a few references are given here, see the ``bibliographical notes'' below.
We encourage interested readers to consult the reviews \cite{FZ,GM,JL}.

\section{Mathematical setting}

\subsection{Definitions}

Let $\Lambda \subset \bbZ^d$ denote a finite cubic box. The configuration space for the classical particles is
$$
\Omega_\Lambda =  \{0,1\}^\Lambda = \big\{ \omega=(\omega_x) : x\in\Lambda, \text{ and } \omega_x = 0,1 \big\},
$$
where $\omega_x = 0$ or 1 denotes the absence or presence of a classical particle at the
site $x$. The total number of classical particles is $\Nc(\omega) = \sum_{x\in\Lambda}
\omega_x$. The Hilbert space for the spinless quantum particles (``electrons'') is the usual fermionic Fock space
$$
\caF_\Lambda = \bigoplus_{N=0}^{|\Lambda|} \caH_{\Lambda,N},
$$
where $\caH_{\Lambda,N}$ is the Hilbert space of square summable, antisymmetric, complex functions $\Psi = \Psi(x_1,\dots,x_N)$ of $N$ variables $x_i \in \Lambda$. Let $a^\dagger_x$ and $a_x$ denote the standard creation and annihilation operators of an electron at $x$; recall that they satisfy the anticommutation relations
$$
\{a_x,a_y\}=0, \quad \{a^\dagger_x,a^\dagger_y\}=0, \quad \{a_x,a^\dagger_y\} = \delta_{xy}.
$$
The Hamiltonian for the Falicov-Kimball model is an operator on $\caF_\Lambda$ that depends on the configurations of classical particles. Namely, for $\omega \in \Omega_\Lambda$, we define
$$
H_\Lambda(\omega) = -\sumtwo{x,y\in\Lambda}{|x-y|=1} a^\dagger_x a_y - U \sum_{x\in\Lambda} \omega_x a^\dagger_x a_x.
$$
The first term represents the kinetic energy of the electrons. The second term represents the on-site attraction ($U>0$) or repulsion ($U<0$) between electrons and classical particles.

The Falicov-Kimball Hamiltonian can be written with the help of a one-body Hamiltonian $h_\Lambda$, which is an operator on the Hilbert space for a single electron $\ell^2(\Lambda)$. Indeed, we have
$$
H_\Lambda(\omega) = \sum_{x,y\in\Lambda} h_{xy}(\omega) a^\dagger_x a_y.
$$
The matrix $h_\Lambda(\omega) = (h_{xy}(\omega))$ is the sum of a hopping matrix (adjacency matrix) $t_\Lambda$, and of a matrix $v_\Lambda(\omega)$ that represents an external potential due to the classical particles. Namely, we have
$$
h_{xy}(\omega) = -t_{xy} - U \omega_x \delta_{xy},
$$
where $t_{xy}$ is one if $x$ and $y$ are nearest-neighbors, and is zero otherwise. The
spectrum of $t_\Lambda$ lies in $(-2d,2d)$, and the eigenvalues of $v_\Lambda(\omega)$ are
$-U$ (with degeneracy $\Nc(\omega)$) and 0 (with degeneracy $|\Lambda|-\Nc(\omega)$).
Denoting $\lambda_j(A)$ the eigenvalues of a matrix $A$, it follows from the minimax principle that:
$$
\lambda_j(A) - \|B\| \leq \lambda_j(A+B) \leq \lambda_j(A) + \|B\|.
$$
Let $\lambda_1(\omega) \leq \lambda_2(\omega) \leq \dots \leq \lambda_{|\Lambda|}(\omega)$ be the eigenvalues of $h_\Lambda(\omega)$.
Choosing $A = v_\Lambda(\omega)$ and $B = t_\Lambda$ in the inequality above, we find that
for $U>0$,
\[
\begin{split}
-U-2d < \lambda_j(\omega) &< -U+2d \quad \text{for } j=1,\dots,\Nc(\omega),\\
-2d < \lambda_j(\omega) &< 2d \hspace{14.5mm} \text{for } j=\Nc(\omega)+1,\dots,|\Lambda|.
\end{split}
\]
In particular, for any configuration $\omega$ and any $\Lambda$,
$$
{\rm Spec} \; h_\Lambda(\omega) \subset (-U-2d,-U+2d) \cup (-2d,2d).
$$
Thus for $U > 4d$ the spectrum of $h_\Lambda(\omega)$ has the ``universal'' gap
$(-U+2d,-2d)$. A similar property holds for $U<-4d$.

\subsection{Canonical ensemble}
\label{seccanens}

A fruitful approach towards understanding the behavior of the Falicov-Kimball model is to
fix first the configuration of the classical particles and to introduce the ground state
energy $E_\Lambda(\Ne,\omega)$ as the lowest eigenvalue of $H_\Lambda(\omega)$ in the subspace $\caH_{\Lambda,\Ne}$:
$$
E_\Lambda(\Ne,\omega) = \inf_{\Psi \in \caH_{\Lambda,\Ne}, \|\Psi\|=1} \bra\Psi
H_\Lambda(\omega) \ket\Psi = \sum_{j=1}^{\Ne} \lambda_j(\omega).
$$
A typical problem is to find the set of  ground state configurations, i.e.\ the set of configurations that minimize
$E_\Lambda(\Ne,\omega)$ for given $\Ne$ and $\Nc=\Nc(\omega)$.

In the case $U>4d$ and $\Ne=\Nc(\omega)$, the ground state energy
$E_\Lambda(\Nc(\omega),\omega)$ has a convergent expansion in powers of $U^{-1}$:
\be
E_\Lambda(\Nc(\omega),\omega) = -U\Nc(\omega) + \sum_{k\geq2} \frac1{k\,U^{k-1}} \sumthree{x_1,\dots,x_k \in \Lambda}{|x_i-x_{i-1}|=1}{0<m(\{x_i\})<k} (-1)^{m(\{x_i\})} \left( \begin{matrix} k-2 \\ m(\{x_i\})-1 \end{matrix} \right),
\label{expgsen}
\ee
where $m(x_1,\dots,x_k)$ is the number of sites $x_i$ with $\omega_{x_i}=0$. The last sum
also includes the condition $|x_k-x_1|=1$. Simple estimates show that the series is less than $\frac{2d}{U-4d}
\Nc(\omega)$. The lowest order term is a nearest-neighbor interaction,
$$
-\frac1U \sum_{\{x,y\}: |x-y|=1} \delta_{w_x,1-w_y},
$$
that favors pairs with different occupation numbers. Formula \eqref{expgsen} is the
starting point for most studies of the phase diagram for large $U$. A similar expansion
holds for $U<-4d$ and $\Ne=|\Lambda|-\Nc(\omega)$.

Phase diagrams are better discussed in the limit of infinite volumes where boundary
effects can be discarded. Let $\Omega^{\rm per}$ be the set of configurations on $\bbZ^d$ that are periodic
in all $d$ directions, and $\Omega^{\rm per}(\rhoc) \subset \Omega^{\rm per}$ be the set
of periodic configurations with density $\rhoc$. For $\omega \in \Omega^{\rm per}$ and
$\rhoe \in [0,1]$, we introduce the energy per site in the infinite volume limit by
\be
\label{ecan}
e(\rhoe,\omega) = \lim_{\Lambda\nearrow\bbZ^d} \tfrac1{|\Lambda|} E_\Lambda(\Ne,\omega).
\ee
Here, the limit is taken over any sequence of increasing cubes, and $\Ne =
\lowerint{\rhoe|\Lambda|}$ is the integer part of $\rhoe|\Lambda|$.
Existence of this limit follows from standard arguments.

In the case of the empty configuration $\omega_x \equiv 0$, we get the
well-known energy per site of free lattice electrons: For $k \in [-\pi,\pi]^d$, let
$\varepsilon(k) = -\sum_{\nu=1}^d \cos k_\nu$; then
$$
e(\rhoe,\omega\equiv0) = \frac1{(2\pi)^d} \int_{\varepsilon(k) < \varepsilon_{\rm
F}(\rhoe)} \varepsilon(k) \dd k,
$$
where $\varepsilon_{\rm F}(\rhoe)$ is the Fermi energy, defined by
$$
\rhoe = \frac1{(2\pi)^d} \int_{\varepsilon(k) < \varepsilon_{\rm F}(\rhoe)} \dd k.
$$
The other simple situation is the full configuration $\omega_x \equiv 1$, whose
energy is $e(\rhoe,\omega\equiv1) = e(\rhoe,\omega\equiv0) - U\rhoe$.

Let $e(\rhoe,\rhoc)$ denote the absolute ground state energy density, namely,
$$
e(\rhoe,\rhoc) = \inf_{\omega\in\Omega^{\rm per}(\rhoc)} e(\rhoe,\omega).
$$
Notice that $e(\rhoe,\omega)$ is convex in $\rhoe$, and that $e(\rhoe,\rhoc)$
is the convex envelope of $\{ e(\rhoe,\omega) : \omega\in\Omega^{\rm per}(\rhoc) \}$. It may be locally
linear around some $(\rhoe,\rhoc)$. This is the case if the infimum is not realized by a
periodic configuration. The non-periodic ground states can be expressed as linear
combinations of two or more periodic ground states (``mixtures''). That is, for $1\leq
i\leq n$ there are $\alpha_i\geq0$ with $\sum_i \alpha_i=1$, $\omega^{(i)} \in \Omega^{\rm
per}$, and $\rhoe^{(i)}$, such that
$$
\rhoe = \sum_i \alpha_i \rhoe^{(i)}, \quad\quad \rhoc = \sum_i \alpha_i
\rhoc(\omega^{(i)}),
$$
and
$$
e(\rhoe,\rhoc) = \sum_i \alpha_i e(\rhoe^{(i)},\omega^{(i)}).
$$
The simplest mixture is the ``segregated state'' for densities $\rhoe<\rhoc$: Take $\omega^{(1)}$
to be the empty configuration, $\omega^{(2)}$ to be
the full configuration, $\rhoe^{(1)}=0$, $\rhoe^{(2)}=\frac\rhoe\rhoc$, and $\alpha_2 = 1-\alpha_1 = \rhoc$.

If $d\geq2$, a mixture between configurations $\omega^{(i)}$ can be realized as follows.
First, partition $\bbZ^d$ into
domains $D_1 \cup \dots \cup D_n$ such that $\frac{|D_i|}{|\Lambda|} \to \alpha_i$ and
$\frac{|\partial D_i|}{|\Lambda|} \to 0$ as $\Lambda \nearrow \bbZ^d$. Then define a
non-periodic configuration $\omega$ by setting
$\omega_x=\omega^{(i)}_x$ for $x\in D_i$. See the
illustration in Fig.\ \ref{figmixedconf}. The canonical energy can be computed from
\eqref{ecan}, and it is equal to
$$
e(\rhoe,\omega) = \inf_{(\rho_{\rm e}^{(i)}): \sum_i \alpha_i \rho_{\rm e}^{(i)} = \rhoe}
\sum_{i=1}^n \alpha_i \; e(\rho_{\rm e}^{(i)},\omega^{(i)}).
$$
Furthermore, the infimum is realized by densities $\rhoe^{(i)}$ such that there exists $\mue$ with
$\rhoe(\mue,\omega^{(i)}) = \rhoe^{(i)}$ for all $i$. (See \eqref{defdenschempot} below
for the definition of $\rhoe(\mue,\omega)$.)

\bfig
\epsfxsize=60mm
\centerline{\epsffile{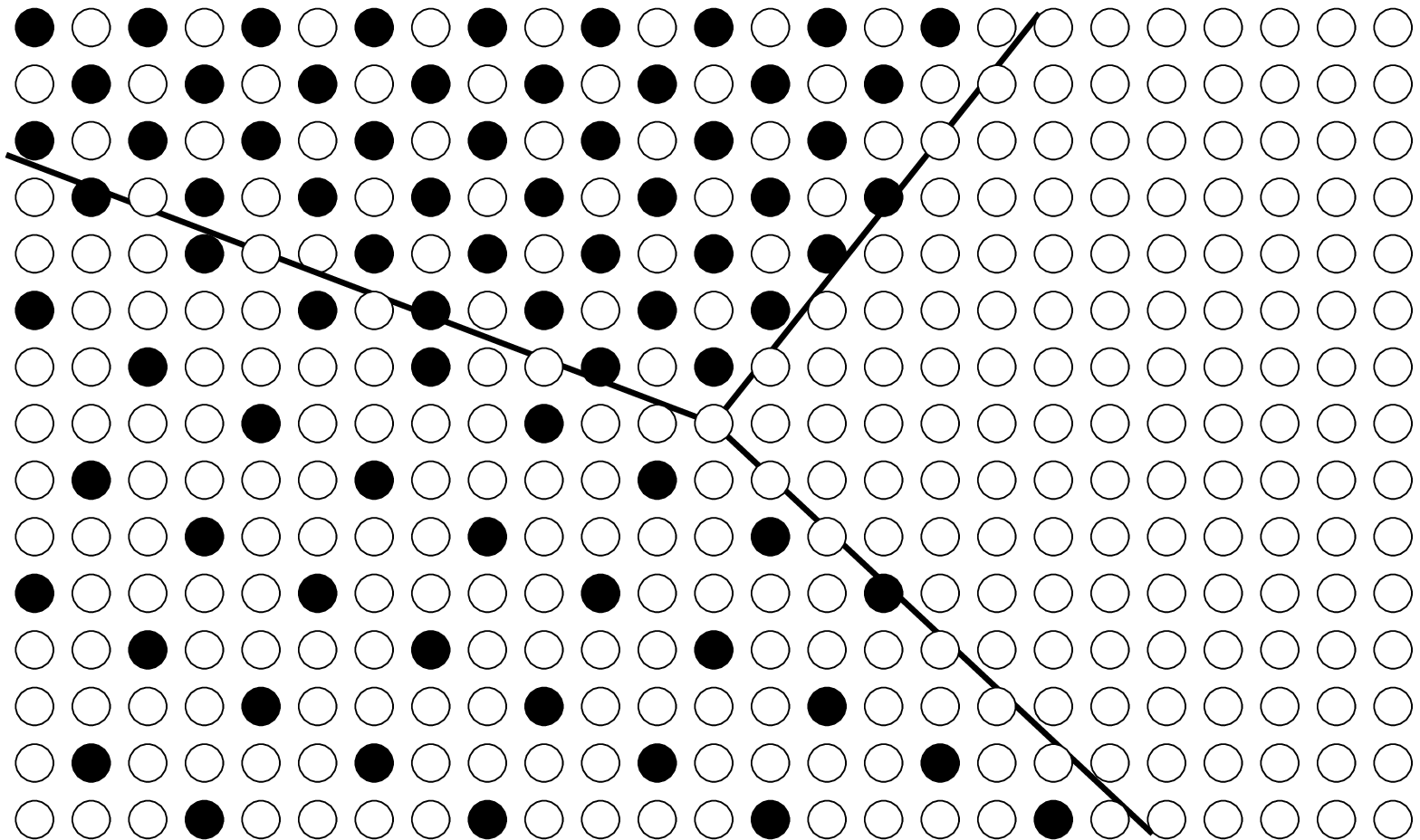}}
\caption{A two-dimensional mixed configuration formed by periodic configurations of
densities 0, $\frac15$, and $\frac12$.}
\label{figmixedconf}
\end{figure}

We define the {\it canonical ground state phase diagram} as the set of ground states
$\omega$ (either a periodic configuration or a mixture) that minimize the ground state
energy for given densities $\rhoe,\rhoc$:
$$
G_{\rm can}(\rhoe,\rhoc) = \bigl\{ \omega :e(\rhoe,\omega) =
e(\rhoe,\rhoc) \text{ and } \rhoc(\omega) = \rhoc \bigr\}.
$$

\subsection{Grand-canonical ensemble}

Properties of the system at finite temperatures are usually investigated within the grand-canonical formalism. The equilibrium state is characterized by an inverse temperature $\beta = 1/k_{\rm B} T$, and by chemical potentials $\mue, \muc$, for the electrons and for the classical particles respectively. In this formalism the thermodynamic properties are derived from the partition functions
\be
\label{partfcts}
\begin{split}
Z_\Lambda(\beta,\mue,\omega) &=  \Tr_{\caF_\Lambda} \e{-\beta [H_\Lambda(\omega) - \mue N_\Lambda]}, \\
Z_\Lambda(\beta,\mue,\muc) &=  \sum_{\omega\in\Omega_\Lambda} \e{\beta\muc\Nc(\omega)} Z_\Lambda(\beta,\mue,\omega).
\end{split}
\ee
Here, $N_\Lambda = \sum_{x\in\Lambda} a^\dagger_x a_x$ is the operator for the total number of electrons. We then define the free energy by
$$
F_\Lambda(\beta,\mue,\muc) = -\tfrac1\beta \log Z_\Lambda(\beta,\mue,\muc).
$$
The first partition function in \eqref{partfcts} allows to introduce an effective
interaction for the classical particles, mediated by the electrons, by
$$
F_\Lambda(\beta,\mue,\muc,\omega) = -\muc\Nc(\omega) - \tfrac1\beta \log Z_\Lambda(\beta,\mue,\omega).
$$
It depends on the inverse temperature $\beta$.
Taking the limit of zero temperature gives the corresponding ground state energy of the
electrons in the classical configuration $\omega$,
$$
E_\Lambda(\mue,\muc,\omega) = \lim_{\beta\to\infty} F_\Lambda(\beta,\mue,\muc,\omega) = -\muc\Nc(\omega) + \sum_{j: \lambda_j(\omega) < \mue} (\lambda_j(\omega) - \mue).
$$
Notice that $F_\Lambda$ and $E_\Lambda$ are strictly decreasing and concave in $\mue,\muc$ ($E_\Lambda$ is
actually linear in $\muc$). We also define the energy density in the infinite volume limit
by considering a sequence of increasing cubes. For $\omega \in \Omega^{\rm per}$,
$$
e(\mue,\muc,\omega) = \lim_{\Lambda\nearrow\bbZ^d} \tfrac1{|\Lambda|}
E_\Lambda(\mue,\muc,\omega).
$$
The corresponding electronic density is
\be
\label{defdenschempot}
\rhoe(\mue,\omega) = \lim_{\Lambda\nearrow\bbZ^d} \tfrac1{|\Lambda|} \,\#\{ j:
\lambda_j(\omega) < \mue \} = -\tfrac\partial{\partial\mue} e(\mue,\muc,\omega),
\ee
and the density of classical particles is $\rhoc(\omega) = \lim_\Lambda
\Nc(\omega)/|\Lambda|$. One can check that canonical and grand-canonical energies are
related by
\be
\label{ecangdcan}
e(\mue,\muc,\omega) = e(\rhoe(\mue,\omega),\omega) - \mue \rhoe(\mue,\omega) - \muc
\rhoc(\omega).
\ee

Given $(\mue,\muc)$, the ground state energy density $e(\mue,\muc)$ is defined by
$$
e(\mue,\muc) = \inf_{\omega\in\Omega^{\rm per}} e(\mue,\muc,\omega).
$$
The set of periodic ground state configurations for given chemical
potentials $\mue,\muc$ is the {\it grand-canonical ground state phase diagram}:
$$
G_{\rm gc}(\mue,\muc) = \bigl\{ \omega \in \Omega^{\rm per} : e(\mue,\muc,\omega) = e(\mue,\muc) \bigr\}.
$$
It may happen that no periodic configuration minimizes
$e(\mue,\muc,\omega)$ and that $G_{\rm gc}(\mue,\muc) = \emptyset$. Results suggest that
$G_{\rm gc}(\mue,\muc)$ is nonempty for almost all $\mue,\muc$, however.

The situation simplifies for $U>4d$ and $\mue \in (-U+2d,-2d)$. Since $\mue$ belongs to the gap of
$h_\Lambda(\omega)$, we have $\rhoe(\mue,\omega)=\rhoc(\omega)$, and
$$
e(\mue,\muc,\omega) = e(\rhoc(\omega),\omega) - (\mue+\muc) \rhoc(\omega).
$$
Thus $G_{\rm gc}(\mue,\muc)$ is invariant along the line $\mue+\muc=\const$ (for $\mue$ in
the gap).

\subsection{Symmetries of the model}

The Hamiltonian $H_\Lambda$ clearly has the symmetries of the lattice (for a box with
periodic boundary conditions, there is invariance under translations, rotations by
90$^\circ$, and reflections through an axis). More important, it also possesses
particle-hole symmetries and these are useful since they allow to restrict investigations to positive $U$ and to certain domains of densities or chemical potentials (see below).
\begin{itemize}
\item The classical particle--hole transformation $\omega_x \mapsto \overline \omega_x = 1-\omega_x$ results in
$$
H_\Lambda^U(\overline \omega) = H_\Lambda^{-U}(\omega) - U N_\Lambda,
$$
and $\Nc(\overline \omega) = |\Lambda| - \Nc(\omega)$. It follows that $E_\Lambda^U(\Ne,\overline \omega) = E_\Lambda^{-U}(\Ne,\omega) - U\Ne$, and
\[
\begin{split}
&G^{-U}_{\rm can}(\rhoe,\rhoc) = \bigl\{ \overline \omega : \omega \in
G^U_{\rm can}(\rhoe,1-\rhoc) \bigr\}, \\
&G^{-U}_{\rm gc}(\mue,\muc) = \bigl\{ \overline \omega : \omega \in
G^U_{\rm gc}(\mue-U,-\muc) \bigr\}. \nn
\end{split}
\]
\item An electron--hole transformation can be defined via the unitary transformation $a_x \mapsto \varepsilon_x a_x^\dagger$ and $a_x^\dagger \mapsto \varepsilon_x a_x$, where $\varepsilon_x$ is equal to 1 on a sublattice, and to $-1$ on the other sublattice. Then
$$
H_\Lambda^U(\omega) \mapsto H_\Lambda^{-U}(\omega) - U\Nc(\omega),
$$
and $N_\Lambda \mapsto |\Lambda|-N_\Lambda$. It follows that $E_\Lambda^U(|\Lambda|-\Ne,\omega) = E_\Lambda^{-U}(\Ne,\omega) - U\Nc(\omega)$, and
\[
\begin{split}
&G^{-U}_{\rm can}(\rhoe,\rhoc) = G^U_{\rm can}(1-\rhoe,\rhoc), \\
&G^{-U}_{\rm gc}(\mue,\muc) = G^U_{\rm gc}(-\mue,\muc-U). \nn
\end{split}
\]
\item Finally, the particle--hole transformation for both the classical particles and the electrons give
$$
H_\Lambda^U(\overline \omega) \mapsto H_\Lambda^U(\omega) + U N_\Lambda + U\Nc(\omega) - U|\Lambda|.
$$
It follows that $E_\Lambda^U(|\Lambda|-\Ne,\overline \omega) = E_\Lambda^U(\Ne,\omega) + U(\Ne+\Nc(\omega)-|\Lambda|)$, and \[
\begin{split}
&G^U_{\rm can}(\rhoe,\rhoc) = \bigl\{ \overline \omega : \omega \in
G^U_{\rm can}(1-\rhoe,1-\rhoc) \bigr\}, \\
&G^U_{\rm gc}(\mue,\muc) = \bigl\{ \overline \omega : \omega \in
G^U_{\rm gc}(-\mue-U,-\muc-U) \bigr\}. \nn
\end{split}
\]
\end{itemize}

Any of the first two symmetries allow to choose the sign of $U$. We assume from now on
that $U\geq0$. The third symmetry indicates that the phase diagrams have a point of
central symmetry, given by $\rhoe=\rhoc=\frac12$ in the canonical ensemble and
$\mue=\muc=-\frac U2$ in the grand-canonical ensemble. Consequently, it is enough to study
densities satisfying $\rhoe\leq\frac12$ and chemical potentials satisfying $\mue \leq
-\frac U2$.

These symmetries have also useful consequences at positive temperatures. In particular,
both species of particles have average density $\frac12$ at $\mue=\muc=-\frac U2$, for all
$\beta$.

\section{The ground state --- Arbitrary dimensions}
\label{secalldim}

\subsection{The segregated state}

What follows is best understood in the limit $U\to\infty$ and when $\rhoe<\rhoc$. In this
case the electrons become localized in the domain $\caD_\Lambda(\omega) = \{ x\in\Lambda : w_x=1 \}$
and their energy per site is that of the full configuration, $e(\rho,\omega\equiv1)$ (see
Section \ref{seccanens}), where $\rho=\rhoe/\rhoc$ is the effective electronic density.
The presence of a boundary for $\caD_\Lambda(\omega)$ raises the energy and the correction is roughly proportional to
$$
B_\Lambda(\omega) = \#\bigl\{ (x,y) : x \in \caD_\Lambda(\omega) \text{ and } y \in \bbZ^d
\setminus\caD_\Lambda(\omega) \bigr\}.
$$

\begin{theorem}\hfill
\label{thmseg}
\begin{itemize}
\item[(a)]
Let $\Lambda \subset \bbZ^d$ be a finite box, and $U>4d$. Then for all
$\omega\in\Omega_\Lambda$, and all $\Ne\leq\Nc(\omega)=\Nc$, we have the following upper and lower bounds:
$$
\tfrac1{2d} |e(\tfrac\Ne\Nc)| B_\Lambda(\omega) \geq E_\Lambda(\Ne,\omega) - \Nc
e(\tfrac\Ne\Nc,\omega\equiv1) \geq \Bigl[ a(\tfrac\Ne\Nc) - \gamma(U) \Bigr] B_\Lambda(\omega).
$$
Here, $a(\rho) = a(1-\rho)$ is strictly positive for $0<\rho<1$. $\gamma(U)$ behaves as
$\frac{8d^2}U$ for large $U$, in the sense that $U\gamma(U) \to 8d^2$ as $U\to\infty$.
\item[(b)] For any $\rhoe\neq\rhoc$ that differ from zero, the segregated state is the
unique ground state if $a(\frac\rhoe\rhoc) > \gamma(U)$, i.e.\ if $U$ is large
enough.
\end{itemize}
\end{theorem}

The proof of (a) is rather lengthy and we only show here that it implies (b). Let
$b(\omega) = \lim_\Lambda \frac{B_\Lambda(\omega)}{|\Lambda|}$, and notice that
$b(\omega)=0$ for the empty, the full, and the segregated configurations;
$0<b(\omega)<d$ for all other periodic configurations or mixtures. Recall that $\rhoc
e(\frac\rhoe\rhoc,\omega\equiv1)$ is the energy density of the segregated state. For all
densities such that $a(\frac\rhoe\rhoc) > \gamma(U)$, and all configurations such that
$\rhoc(\omega)=\rhoc$, we have
$$
e(\rhoe,\omega) \geq \rhoc \, e(\tfrac\rhoe\rhoc,\omega\equiv1),
$$
and the inequality is strict for any periodic configuration. This shows that the
segregated configuration is the unique ground state.

\subsection{General properties of the grand-canonical phase diagram}
\label{gcphd}

We have already seen that the grand-canonical phase diagram is symmetric with respect to
$(-\frac U2,-\frac U2)$. Other properties follow from concavity of $e(\mue,\muc)$.

Let $\omega \in G_{\rm gc}(\mue,\muc) \setminus G_{\rm gc}(\mue',\muc')$ and $\omega' \in G_{\rm gc}(\mue',\muc')
\setminus G_{\rm gc}(\mue,\muc)$. Then
\begin{itemize}
\item[(i)] $\mue=\mue'$ and $\muc'>\muc$ imply $\rhoc(\omega') > \rhoc(\omega)$;
\item[(ii)] $\muc=\muc'$ and $\mue'>\mue$ imply $\rhoe(\mue',\omega') >
\rhoe(\mue,\omega)$, and $\omega$ cannot be obtained by adding some classical particles to
the configuration $\omega'$.
\end{itemize}
It follows from (ii) that if $\omega\equiv1 \in G_{\rm gc}(\mue,\muc)$, then
$\omega\equiv1 \in G_{\rm gc}(\mue',\muc')$ for all $\mue\geq\mue'$, $\muc\geq\muc'$. A
similar property holds for the empty configuration. To establish these properties, we can
start from
\be
\label{uneinegalite}
e(\mue,\muc,\omega') - e(\mue,\muc,\omega) > 0 > e(\mue',\muc',\omega') -
e(\mue',\muc',\omega).
\ee
Since $e(\mue,\muc,\omega)$ is concave with respect to $\mue$ and linear with respect to
$\muc$, we have
\be
\label{concavity}
e(\mue',\muc',\omega) \leq e(\mue,\muc,\omega) + (\mue-\mue') \rhoe(\mue,\omega) +
(\muc-\muc') \rhoc(\omega).
\ee
Using this inequality for both terms of the right side of \eqref{uneinegalite}, we obtain the
inequality
$$
(\mue'-\mue) \bigl[ \rhoe(\mue',\omega') - \rhoe(\mue,\omega) \bigr] + (\muc'-\muc) \bigl[
\rhoc(\omega') - \rhoc(\omega) \bigr] \geq 0,
$$
which proves (i) and the first part of (ii). The second part of (ii) follows from
$$
e(\mue,\muc,\omega) = -\int_{-\infty}^{\mue} \rhoe(\mu,\omega) \dd\mu - \muc \rhoc(\omega).
$$
Indeed, the minimax principle implies that eigenvalues $\lambda_j(\omega)$ are
decreasing with respect to $\omega$ (if $U\geq0$), so that $\rhoe(\mue,\omega)$ is
increasing (with respect to $\omega$). Then for any $\omega''>\omega$ and $\mue'>\mue$,
$$
e(\mue',\muc,\omega'') - e(\mue',\muc,\omega) > e(\mue,\muc,\omega'') - e(\mue,\muc,\omega),
$$
and $\omega'' \notin G_{\rm gc}(\mue,\muc)$ implies $\omega'' \notin G_{\rm
gc}(\mue',\muc)$.

Next we discuss domains in the plane of chemical potentials where the empty, full, and
chessboard configurations have minimum energy. One easily sees that $\omega\equiv1$ is the
unique ground state configuration if $\muc>0$, or if $\mue>2d$ and $\muc>-U$. Similarly,
$\omega\equiv0$ is the unique ground state if $\muc<-U$, or if $\mue<-U-2d$ and $\muc<0$.
For $U>4d$, it follows from the expansion \eqref{expgsen} that the full configuration is also ground state if
$-U+2d<\mue<-2d$ and $\mue+\muc+U > \frac{4d}{U-4d}$. These domains can be rigorously extended using
energy estimates that involve correlation functions of classical particles. The results
are illustrated in Figs\ \ref{figphdgc1} ($U<4d$) and \ref{figphdgc2} ($U>4d$).
\bfig
\epsfxsize=80mm
\centerline{\epsffile{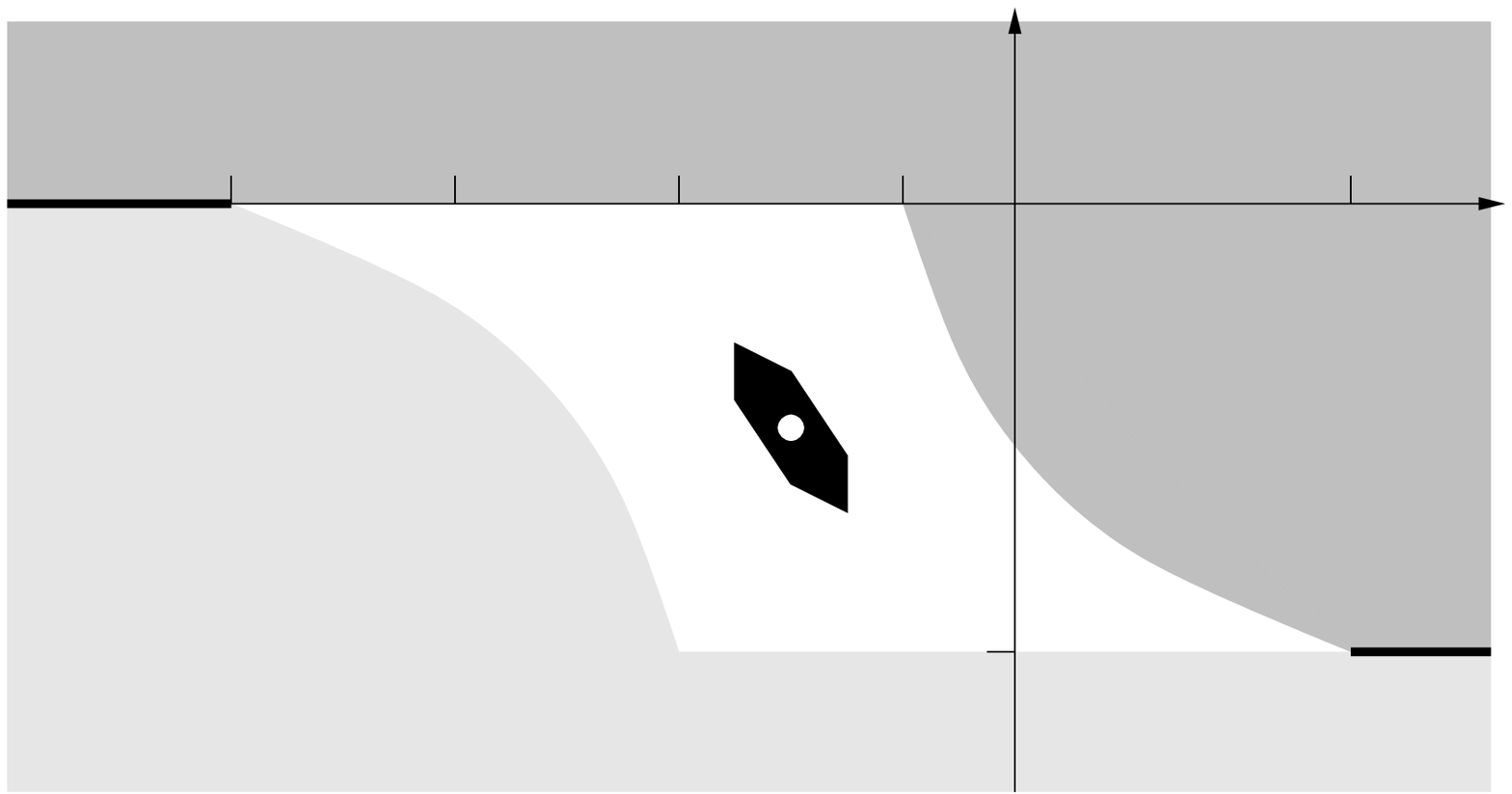}}
\figtext{
\writefig	11.5	3.55	{$\mue$}
\writefig	8.5	4.8	{$\muc$}
\writefig	3.9	3.8	{\tiny $-U-2d$}
\writefig	5.5	3.8	{\tiny $-U$}
\writefig	6.6	3.8	{\tiny $-2d$}
\writefig	7.5	3.8	{\tiny $-U+2d$}
\writefig	10.35	3.8	{\tiny $2d$}
\writefig	8.1	1.15	{\tiny $-U$}
\writefig	4.2	1.6	{$\omega\equiv0$}
\writefig	9.8	2.5	{$\omega\equiv1$}
\writefig	6.1	2.0	{\tiny $(-\tfrac U2, -\tfrac U2)$}
}
\caption{Grand-canonical ground state phase diagram for $U<4d$. Domains for the empty,
chessboard, and full configurations, are denoted in light gray, black, and dark gray
respectively.}
\label{figphdgc1}
\end{figure}

\bfig
\epsfxsize=80mm
\centerline{\epsffile{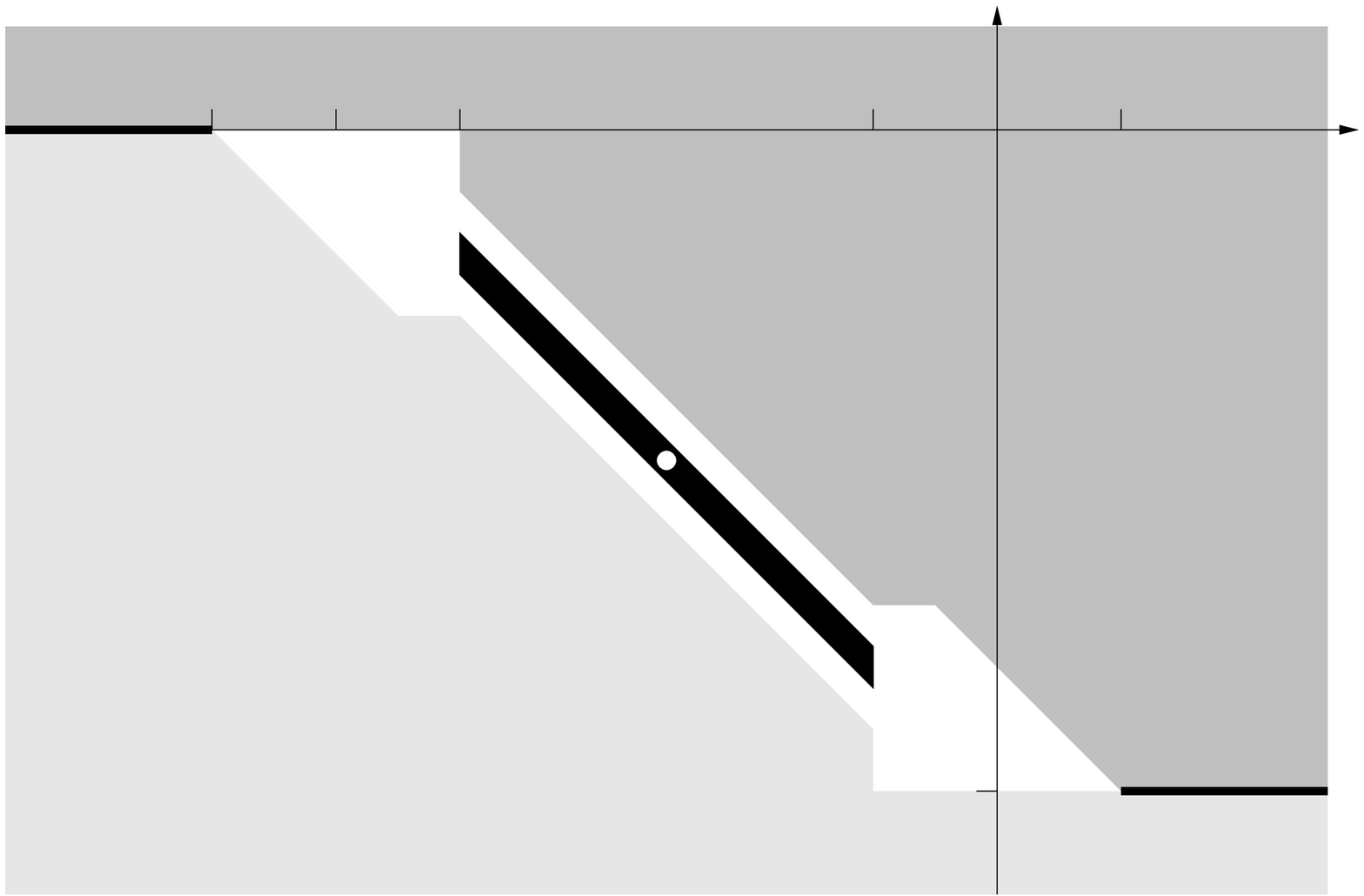}}
\figtext{
\writefig	11.5	4.9	{$\mue$}
\writefig	9.0	5.9	{$\muc$}
\writefig	5.05	5.4	{\footnotesize $-U$}
\writefig	4.0	5.15	{\tiny $-U-2d$}
\writefig	5.55	5.15	{\tiny $-U+2d$}
\writefig	8.15	5.15	{\footnotesize $-2d$}
\writefig	9.8	5.15	{\footnotesize $2d$}
\writefig	8.5	1.0	{\footnotesize $-U$}
\writefig	4.8	1.5	{$\omega\equiv0$}
\writefig	7.8	4.0	{$\omega\equiv1$}
\writefig	5.7	2.7	{\footnotesize $(-\tfrac U2, -\tfrac U2)$}
}
\caption{Grand-canonical ground state phase diagram for $U>4d$. Domains for the empty,
chessboard, and full configurations, are denoted in light gray, black, and dark gray
respectively.}
\label{figphdgc2}
\end{figure}

Finally, canonical and grand-canonical phase diagrams are related by the following
properties:
\begin{itemize}
\item[(iii)] If $\omega \in G_{\rm gc}(\mue,\muc)$, then $\omega \in G_{\rm can}(\rhoe(\mue,\omega),\rhoc(\omega))$.
\item[(iv)] More generally, suppose that $\omega^{(1)},\dots,\omega^{(n)} \in G_{\rm
gc}(\mue,\muc)$, and consider a mixture with coefficients $\alpha_1,\dots,\alpha_n$. The
mixture belongs to
$G_{\rm can}(\rhoe,\rhoc)$, with $\rhoe = \sum_i \alpha_i \rhoe(\mue,\omega^{(i)})$ and
$\rhoc = \sum_i \alpha_i \rhoc(\omega^{(i)})$.
\end{itemize}
To establish (iii), observe that any $\omega'$ satisfies $e(\mue,\muc,\omega') \geq
e(\mue,\muc,\omega)$ if $\omega \in G_{\rm gc}(\mue,\muc)$. Let $\rhoe=\rhoe(\mue,\omega)$ and
$\rhoc=\rhoc(\omega)$, and let $\mue'$ such that $\rhoe(\mue',\omega') = \rhoe$. By Eqs
\eqref{ecangdcan} and \eqref{concavity},
$$
e(\rhoe(\mue',\omega'),\omega') - \mue\rhoe(\mue',\omega') - \muc\rhoc(\omega') \geq
e(\rhoe(\mue,\omega),\omega) - \mue\rhoe(\mue,\omega) - \muc\rhoc(\omega).
$$
Then $e(\rhoe,\omega') \geq e(\rhoe,\omega)$ for any configuration $\omega'$ such that
$\rhoc(\omega')=\rhoc$.
Property (iv) follows from (iii) by a limiting argument, because a mixture can be
approximated by a sequence of periodic configurations.

Next we describe further properties of the phase diagrams that are specific to dimensions
1 and 2.

\section{Ground state configurations --- Dimension one}

A large number of investigations, either analytical or numerical, have been devoted to the
study of the ground state configurations in one dimension. One-dimensional results also serve
as guide to higher dimensions. Recall that symmetries allow to restrict to $U\geq0$ and
$\rhoe\leq\frac12$.

Most ground state configurations that appear in the canonical phase diagram seem to be
given by an intriguing formula, which we now describe. Let $\rhoe=\frac pq$ with $p$
relatively prime to $q$. Then corresponding periodic ground state configurations have
period $q$ and density $\rhoc=\frac rq$ ($r$ is an integer). The occupied
sites in the cell $\{0,1,\dots,q-1\}$ are given by the solutions $k_0,\dots,k_{r-1}$ of
\be
\label{intriguingformula}
(p k_j) = j \mod q, \quad\quad 0\leq j\leq r-1.
\ee
Note that the first classical particle is located at $k_0=0$, and
$k_0,\dots,k_{p-1}$ are {\it not} in increasing order. In order to discuss the solutions
of \eqref{intriguingformula}, we introduce $\ell = \lowerint{\frac qp}$ (the integer part
of $\frac qp$), and we write
\be
\label{undetail}
q = (\ell+1)p - s,
\ee
where $1\leq s\leq p-1$, and $s$ is relatively prime to $p$. Next, let $L(x)$ denote the
distance between the particle at $x$ and the one immediately preceding it (to the left).

Let us observe that if $\rhoc=\rhoe$, i.e.\ if $r=p$, then
\begin{itemize}
\item[(a)] $L(k_j)=\ell$ for $0\leq j\leq s-1$ and $k_j-\ell = k_{j+p-s}$.
\item[(b)] $L(k_j)=\ell+1$ for $s\leq j\leq p-1$ and $k_j-(\ell+1) = k_{j-s}$.
\end{itemize}

Indeed, for $pk_j = j+nq$, Eq.\ \eqref{undetail} implies
$$
p(k_j-\ell) = j+ (n-1)q + (p-s) = j+p-s \mod q,
$$
and
$$
p(k_j-\ell-1) = j-s  \mod q.
$$
Therefore $k_j-\ell$ is solution of \eqref{intriguingformula} if $j+p-s\leq p-1$, while
$k_j - (\ell+1)$ is solution of \eqref{intriguingformula} if $j-s\geq0$.

These properties show that the configuration defined by \eqref{intriguingformula} is
such that $L(x) \in \{\ell,\ell+1\}$ for all occupied $x$. A periodic configuration such
that all distances between consecutive particles are either $\ell$ or $\ell+1$ is called
{\it homogeneous}. Let $\omega$ be a homogeneous configuration with period $q$ and density
$\rhoc=\frac rq$, and let $x_0<\dots<x_{p-1}$ be the occupied sites in
$\{0,1,\dots,q-1\}$.
We introduce the {\it derivative} $\omega'$ of $\omega$ as the periodic
configuration with period $r$ defined by (see Fig.\ \ref{figderconf})
\[
\omega_i' = \begin{cases} 1 & \text{if } L(x_i) = \ell, \\ 0 & \text{if }  L(x_i)
= \ell+1. \end{cases}
\]
A configuration is {\it most homogeneous} if it can be ``differentiated'' repeatedly until
the empty or the full configuration is obtained.

\setlength{\unitlength}{5mm}
\begin{figure}[h]
\centering \begin{picture}(26,5)

\put(0,3.8){$\omega$:}
\put(3,4){\circle*{0.5}} \put(4,4){\circle{0.5}} \put(5,4){\circle{0.5}}
\put(6,4){\circle{0.5}} \put(7,4){\circle*{0.5}} \put(8,4){\circle{0.5}}
\put(9,4){\circle{0.5}} \put(10,4){\circle*{0.5}} \put(11,4){\circle{0.5}}
\put(12,4){\circle{0.5}} \put(13,4){\circle{0.5}} \put(14,4){\circle*{0.5}}
\put(15,4){\circle{0.5}} \put(16,4){\circle{0.5}} \put(17,4){\circle*{0.5}}
\put(18,4){\circle{0.5}} \put(19,4){\circle{0.5}} \put(20,4){\circle{0.5}}
\put(21,4){\circle*{0.5}} \put(22,4){\circle{0.5}} \put(23,4){\circle{0.5}}
\put(24,4){\circle*{0.5}} \put(25,4){\circle{0.5}} \put(26,4){\circle{0.5}}

\put(2,2.7){\footnotesize $k_0=0$} \put(9,2.7){\footnotesize $k_1=7$}
\put(16,2.7){\footnotesize $k_2=14$} \put(23,2.7){\footnotesize $k_3=21$}
\put(6,2.7){\footnotesize $k_4=4$} \put(13,2.7){\footnotesize $k_5=11$}
\put(20,2.7){\footnotesize $k_6=18$}

\put(0,0.8){$\omega'$:}
\put(3,1){\circle*{0.5}} \put(7,1){\circle{0.5}} \put(10,1){\circle*{0.5}}
\put(14,1){\circle{0.5}} \put(17,1){\circle*{0.5}} \put(21,1){\circle{0.5}} \put(24,1){\circle*{0.5}}

\put(2,-0.3){\footnotesize $k_0'=0$} \put(9,-0.3){\footnotesize $k_1'=2$}
\put(16,-0.3){\footnotesize $k_2'=4$} \put(23,-0.3){\footnotesize $k_3'=6$}
\put(6,-0.3){\footnotesize $k_4'=1$} \put(13,-0.3){\footnotesize $k_5'=3$}
\put(20,-0.3){\footnotesize $k_6'=5$}

\end{picture}
\caption{The configuration $\omega$ given by the formula \eqref{intriguingformula} with $q=24$ and
$p=7$, and its derivative $\omega'$. Notice that $\ell=3$ and $s=4$.}
\label{figderconf}
\end{figure}
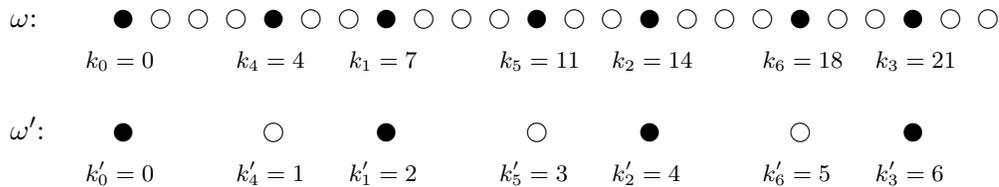

Let $\omega$ be the homogeneous configuration from \eqref{intriguingformula} and $\omega'$
be its derivative. Using the same arguments as for properties (a) and (b) above, and
the fact that $s$ is relatively prime to $p$, we obtain:
\begin{itemize}
\item[(c)] Let $k_0',\dots,k_{p-1}'$ be the solutions of
$$
(s k_j') = j \mod p.
$$
Then $(k_0',\dots,k_{p-1}')$ is a permutation of $(0,1,\dots,p-1)$. Further,
$k_j'-1=k_{j+p-s}'$ for $0\leq j\leq s-1$, and $k_j'-1=k_{j-s}'$ for $s\leq j\leq p-1$.
\end{itemize}

Consider the periodic configuration with period $p$ where sites $k_0',\dots,k_{s-1}'$ are
occupied and sites $k_s',\dots,k_{p-1}'$ are empty. Since $k'_0=0$, this configuration is
precisely the derivative $\omega'$ of $\omega$. Iterating, these properties prove that the
solutions of \eqref{intriguingformula} are most homogeneous.

One of the most important result in one dimension is that only most homogeneous
configurations are present in the canonical phase diagram, for $U$ large enough and for
equal densities $\rhoe=\rhoc$.

\begin{theorem}
\label{thm1d}
Suppose that $\rhoe=\rhoc=\frac pq$. There exists a constant $c$ such that for $U > c
4^q$, the only ground state configuration is the most homogeneous configuration, given by 
\eqref{intriguingformula} (together with translations and reflections).
\end{theorem}

This theorem was established using the expansion \eqref{expgsen} of $E_\Lambda(\Ne,\omega)$ in powers of
$U^{-1}$. It suggests a devil's staircase structure with infinitely many domains.
However, the number of domains for {\it fixed} $U$ could still be finite. Results from
Theorem \ref{thm1d} are illustrated in Fig.\ \ref{figphdgc1d}.

\bfig
\epsfxsize=90mm
\centerline{\epsffile{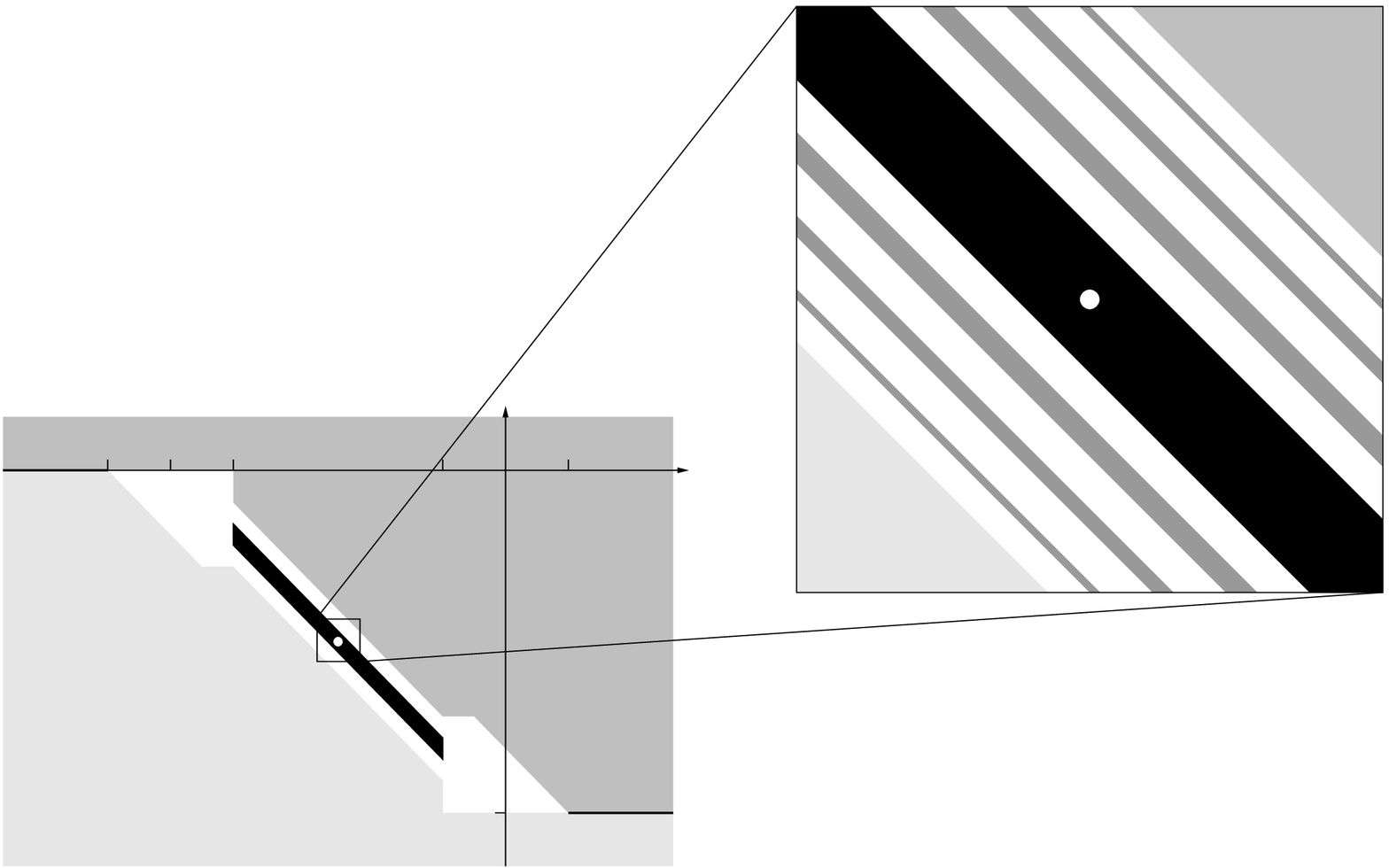}}
\figtext{
\writefig	7.45	3.02	{\tiny $\mue$}
\writefig	6.3	3.55	{\tiny $\muc$}
\writefig	3.7	3.15	{\tiny $-U$}
\writefig	5.65	0.75	{\tiny $-U$}
\writefig	8.2	2.5	{$\omega\equiv0$}
\writefig	10.8	5.6	{$\omega\equiv1$}
\writefig	3.5	1.1	{\footnotesize $\omega\equiv0$}
\writefig	6.3	2.4	{\footnotesize $\omega\equiv1$}
}
\caption{Grand-canonical ground state phase diagram in one dimension for $U>4$ and $\mue$
in the universal gap. Chessboard configurations occur in the black domain. Dark gray
oblique domains correspond to densities $\frac15$, $\frac14$, $\frac13$, $\frac12$,
$\frac23$, $\frac34$, $\frac45$. Total width of these domains is of order $U^{-1}$.}
\label{figphdgc1d}
\end{figure}

For small $U$, on the other hand, one can use a (non-rigorous) Wigner-Brillouin degenerate perturbation
theory (a standard tool in band theory). Let $\rhoe = \frac pq$ with $p$ relatively prime to $q$, and
$\omega$ be a periodic configuration with period $nq$, $n\in\bbN$. Then for $U$ small
enough ($U \ll \frac1q$), we obtain the following expansion for the ground state energy
\be
\label{esmallU}
e(\rhoe,\omega) = -\tfrac2\pi \sin\pi\rhoe - U\rhoe\rhoc(\omega) -
\tfrac{|\widehat\omega(\rhoe)|^2}{4\pi\sin\pi\rhoe} U^2 |\log U| + O(U^2),
\ee
where $\widehat\omega(\rhoe)$ is the ``structure factor'' of the periodic configuration
$\omega$, namely
$$
\widehat\omega(\rhoe) = \frac1{nq} \sum_{j=0}^{nq-1} \e{-2\pi\ii \rhoe j} w_j.
$$
This expansion suggests that the ground state configuration can be found by maximizing the
structure factor. The following theorem holds independently of $U$.

\begin{theorem}
\label{thmstructfact}
Let $\rhoe = \frac pq$. There exist $r_1 \geq \frac q4$ and $r_2 \leq \frac{3q}4$ such that the configurations
maximizing the structure factor are given as follows:
\begin{itemize}
\item[(a)] For $\rhoc = \frac rq$ with $r_1\leq r\leq r_2$, use the formula
\eqref{intriguingformula}.
\item[(b)] For $\rhoc \in (\frac rq, \frac{r+1}q)$ with $r_1\leq r\leq r_2-1$, the
configuration is a mixture of those for $\rhoc = \frac rq$ and $\rhoc = \frac{r+1}q$.
\item[(c)] For $\rhoc \in (0,\frac{r_1}q)$, the configurations are mixtures of $\omega\equiv0$
and that for $\rhoc=\frac{r_1}q$. For $\rhoc \in (\frac{r_2}q,1)$, the configurations are mixtures of $\omega\equiv1$
and that for $\rhoc=\frac{r_2}q$.
\end{itemize}
\end{theorem}

Some insight for low densities is provided by computing the energy of just one classical
particle and one electron on the infinite line, and to compare it with two consecutive
classical particles and two electrons. It turns out that the former is more
favorable than the latter for $U > \frac2{\sqrt3} \approx 1.15$, while ``molecules'' of
two particles are forming when $U < \frac2{\sqrt3}$. Smaller $U$ shows even bigger
molecules for $\rhoc=n\rhoe$, and $n$-molecules are most homogeneously distributed
according to the formula \eqref{intriguingformula}. It should be stressed that the
canonical ground state cannot be periodic if $U$ is small and $\rhoc \notin
[\frac14,\frac34]$, which is different from the case of large $U$.

Only numerical results are available for intermediate $U$. They suggest that
configurations occuring in the phase diagram are essentially given by Theorem
\ref{thmstructfact} (together with the segregated configuration).
This is sketched in Fig.\ \ref{figphdsmallU}, where bold coexistence lines for $\mue>-U-2$
and $\mue<2$ represent segregated states.

\bfig
\epsfxsize=80mm
\centerline{\epsffile{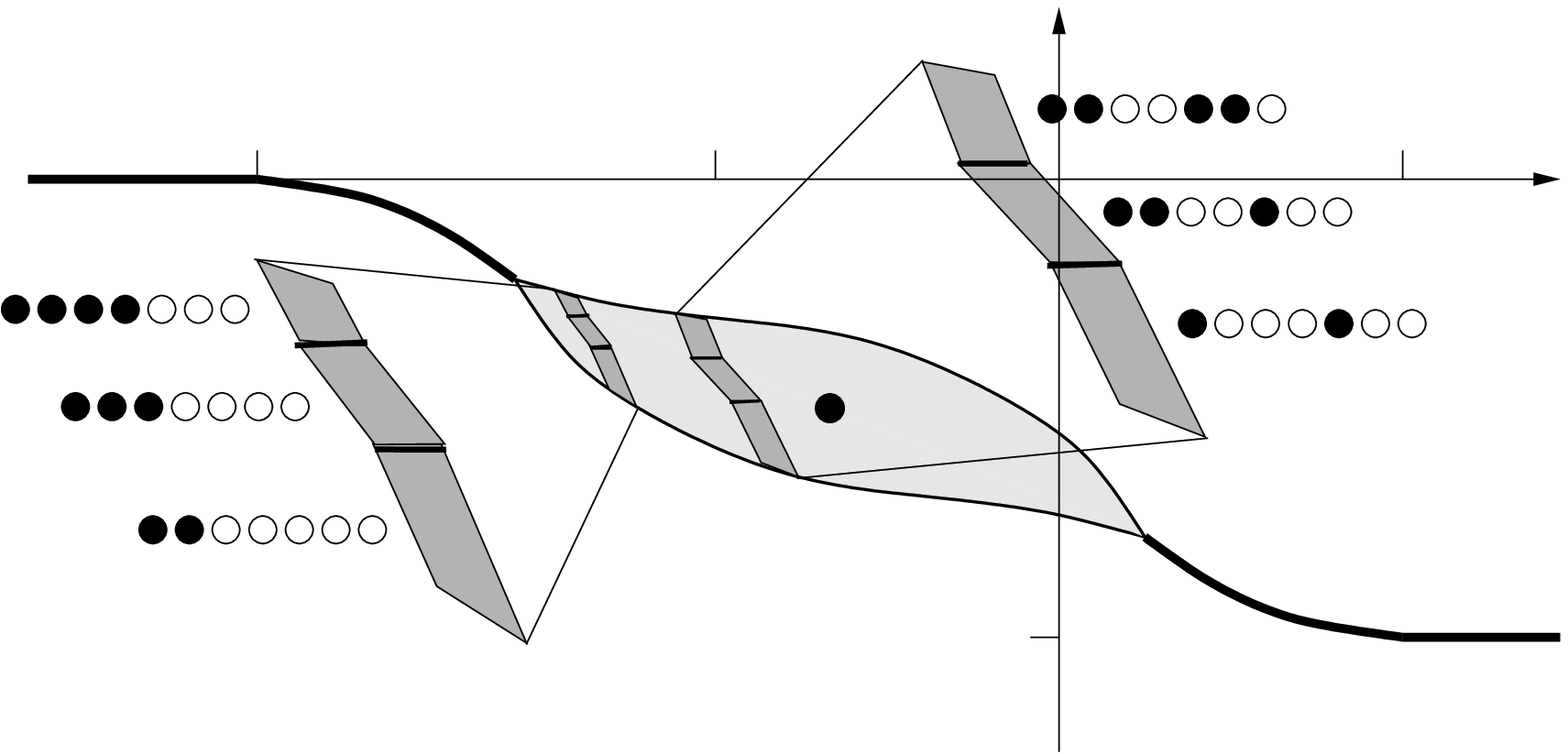}}
\figtext{
\writefig	11.5	3.3	{$\mue$}
\writefig	8.6	4.45	{$\muc$}
\writefig	4.0	3.65	{\footnotesize $-U-2$}
\writefig	6.8	3.65	{\footnotesize $-2$}
\writefig	10.45	3.65	{\footnotesize $2$}
\writefig	8.05	0.95	{\footnotesize $-U$}
\writefig	6.7	0.7	{$\omega\equiv0$}
\writefig	10.1	1.6	{$\omega\equiv1$}
%\writefig	7.45	2.45	{\footnotesize $(-\tfrac U2, -\tfrac U2)$}
}
\caption{Grand-canonical ground state phase diagram for $U\approx0.4$. Enlarged are domains
for $\rhoe=\frac17$ and $\frac27$, with the same densities $\rhoc=\frac27,\frac37,\frac47$.}
\label{figphdsmallU}
\end{figure}

\section{Ground state configurations --- Dimension two}

We discuss the canonical ensemble only, but many results extend to the grand-canonical
ensemble. Recall that $G_{\rm can}(\frac12,\frac12)$ consists of the two chessboard
configurations for any $U>0$, and that segregation takes place when
$\rho_e\neq\rhoc$, providing $U$ is large enough (Theorem \ref{thmseg}). Other results deal
with the case of equal densities, and for $U$ large enough.

\begin{theorem}
\label{thm2d}
Let $\rhoe=\rhoc\equiv\rho\leq\frac12$.
\begin{itemize}
\item[(a)] If $\rho \in \{ \frac12, \frac25, \frac13, \frac14, \frac29, \frac15, \frac2{11},
\frac16 \}$, then for $U$ large enough, the ground state configurations are those
displayed in Fig.\ \ref{figconf}. If $\rho = \frac1{n^2 + (n+1)^2}$ with integer $n$,
then for $U$ large enough (depending on $\rho$), the ground state configurations are
periodic.
\item[(b)] If $\rho$ is a rational number between $\frac13$ and $\frac25$, then for $U$
large enough (depending on the denominator of $\rho$), the ground state configurations are
periodic. Further, the restriction to any horizontal line is a one-dimensional periodic
configuration given by \eqref{intriguingformula}, and the configuration is constant in
either the direction $(\begin{smallmatrix} 1 \\ 1 \end{smallmatrix})$ or
$(\begin{smallmatrix} 1 \\ -1 \end{smallmatrix})$.
\item[(c)] Suppose that $U$ is large enough. If $\rho \in (\frac16,\frac2{11})$, the ground state
configurations are mixtures of the configurations $\rho=\frac16$ and $\rho=\frac2{11}$ of
Fig.\ \ref{figconf}. If $\rho \in (\frac15,\frac29)$, the ground state configurations are mixtures of the configurations $\rho=\frac15$ and
$\rho=\frac29$. If $\rho \in (\frac29,\frac14)$, the ground state configurations are
mixtures of the configurations $\rho=\frac29$ and $\rho=\frac14$.
\end{itemize}
\end{theorem}

The canonical phase diagram for $\rhoe=\rhoc$ is presented in Fig. \ref{figphdcan2d}.

\setlength{\unitlength}{3.6mm}
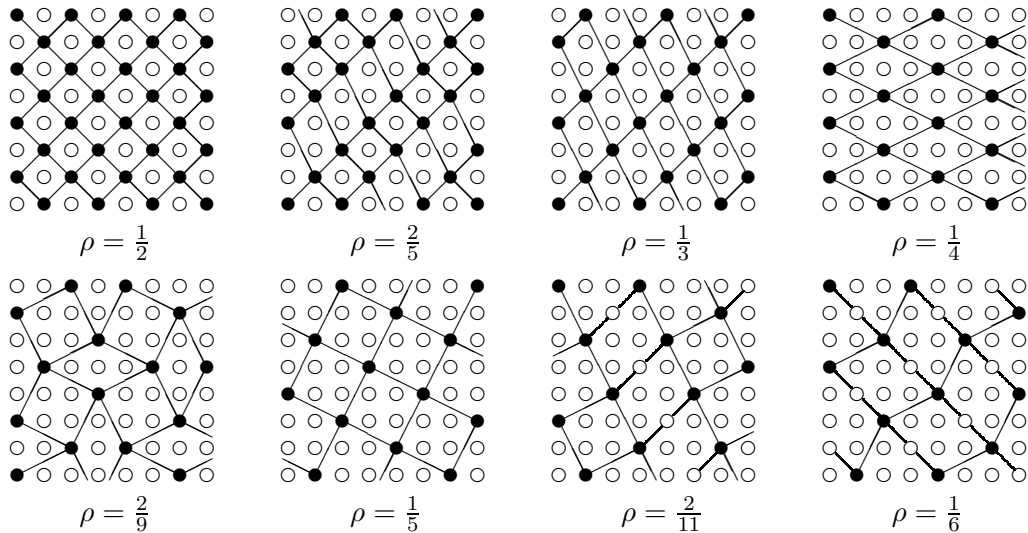
\begin{figure}[h]
\centering \begin{picture}(39,18)
\put(1,11){\circle{0.5}} \put(2,11){\circle*{0.5}} \put(3,11){\circle{0.5}} \put(4,11){\circle*{0.5}} \put(5,11){\circle{0.5}} \put(6,11){\circle*{0.5}} \put(7,11){\circle{0.5}} \put(8,11){\circle*{0.5}} \put(1,12){\circle*{0.5}} \put(2,12){\circle{0.5}} \put(3,12){\circle*{0.5}} \put(4,12){\circle{0.5}} \put(5,12){\circle*{0.5}} \put(6,12){\circle{0.5}} \put(7,12){\circle*{0.5}} \put(8,12){\circle{0.5}} \put(1,13){\circle{0.5}} \put(2,13){\circle*{0.5}} \put(3,13){\circle{0.5}} \put(4,13){\circle*{0.5}} \put(5,13){\circle{0.5}} \put(6,13){\circle*{0.5}} \put(7,13){\circle{0.5}} \put(8,13){\circle*{0.5}} \put(1,14){\circle*{0.5}} \put(2,14){\circle{0.5}} \put(3,14){\circle*{0.5}} \put(4,14){\circle{0.5}} \put(5,14){\circle*{0.5}} \put(6,14){\circle{0.5}} \put(7,14){\circle*{0.5}} \put(8,14){\circle{0.5}} \put(1,15){\circle{0.5}} \put(2,15){\circle*{0.5}} \put(3,15){\circle{0.5}} \put(4,15){\circle*{0.5}} \put(5,15){\circle{0.5}} \put(6,15){\circle*{0.5}} \put(7,15){\circle{0.5}} \put(8,15){\circle*{0.5}} \put(1,16){\circle*{0.5}} \put(2,16){\circle{0.5}} \put(3,16){\circle*{0.5}} \put(4,16){\circle{0.5}} \put(5,16){\circle*{0.5}} \put(6,16){\circle{0.5}} \put(7,16){\circle*{0.5}} \put(8,16){\circle{0.5}} \put(1,17){\circle{0.5}} \put(2,17){\circle*{0.5}} \put(3,17){\circle{0.5}} \put(4,17){\circle*{0.5}} \put(5,17){\circle{0.5}} \put(6,17){\circle*{0.5}} \put(7,17){\circle{0.5}} \put(8,17){\circle*{0.5}} \put(1,18){\circle*{0.5}} \put(2,18){\circle{0.5}} \put(3,18){\circle*{0.5}} \put(4,18){\circle{0.5}} \put(5,18){\circle*{0.5}} \put(6,18){\circle{0.5}} \put(7,18){\circle*{0.5}} \put(8,18){\circle{0.5}}

\put(1,16){\line(1,1){2}} \put(1,14){\line(1,1){4}} \put(1,12){\line(1,1){6}} \put(2,11){\line(1,1){6}} \put(4,11){\line(1,1){4}} \put(6,11){\line(1,1){2}}
\put(1,12){\line(1,-1){1}} \put(1,14){\line(1,-1){3}} \put(1,16){\line(1,-1){5}} \put(1,18){\line(1,-1){7}} \put(3,18){\line(1,-1){5}} \put(5,18){\line(1,-1){3}} \put(7,18){\line(1,-1){1}}

\put(11,11){\circle*{0.5}} \put(12,11){\circle{0.5}} \put(13,11){\circle*{0.5}} \put(14,11){\circle{0.5}} \put(15,11){\circle{0.5}} \put(16,11){\circle*{0.5}} \put(17,11){\circle{0.5}} \put(18,11){\circle*{0.5}} \put(11,12){\circle{0.5}} \put(12,12){\circle*{0.5}} \put(13,12){\circle{0.5}} \put(14,12){\circle*{0.5}} \put(15,12){\circle{0.5}} \put(16,12){\circle{0.5}} \put(17,12){\circle*{0.5}} \put(18,12){\circle{0.5}} \put(11,13){\circle{0.5}} \put(12,13){\circle{0.5}} \put(13,13){\circle*{0.5}} \put(14,13){\circle{0.5}} \put(15,13){\circle*{0.5}} \put(16,13){\circle{0.5}} \put(17,13){\circle{0.5}} \put(18,13){\circle*{0.5}} \put(11,14){\circle*{0.5}} \put(12,14){\circle{0.5}} \put(13,14){\circle{0.5}} \put(14,14){\circle*{0.5}} \put(15,14){\circle{0.5}} \put(16,14){\circle*{0.5}} \put(17,14){\circle{0.5}} \put(18,14){\circle{0.5}} \put(11,15){\circle{0.5}} \put(12,15){\circle*{0.5}} \put(13,15){\circle{0.5}} \put(14,15){\circle{0.5}} \put(15,15){\circle*{0.5}} \put(16,15){\circle{0.5}} \put(17,15){\circle*{0.5}} \put(18,15){\circle{0.5}} \put(11,16){\circle*{0.5}} \put(12,16){\circle{0.5}} \put(13,16){\circle*{0.5}} \put(14,16){\circle{0.5}} \put(15,16){\circle{0.5}} \put(16,16){\circle*{0.5}} \put(17,16){\circle{0.5}} \put(18,16){\circle*{0.5}} \put(11,17){\circle{0.5}} \put(12,17){\circle*{0.5}} \put(13,17){\circle{0.5}} \put(14,17){\circle*{0.5}} \put(15,17){\circle{0.5}} \put(16,17){\circle{0.5}} \put(17,17){\circle*{0.5}} \put(18,17){\circle{0.5}} \put(11,18){\circle{0.5}} \put(12,18){\circle{0.5}} \put(13,18){\circle*{0.5}} \put(14,18){\circle{0.5}} \put(15,18){\circle*{0.5}} \put(16,18){\circle{0.5}} \put(17,18){\circle{0.5}} \put(18,18){\circle*{0.5}}

\put(11,16){\line(1,1){2}} \put(11,14){\line(1,1){4}} \put(11,11){\line(1,1){7}} \put(13,11){\line(1,1){5}} \put(16,11){\line(1,1){2}}
\put(12,17){\line(-1,2){0.6}} \put(17,17){\line(-1,2){0.6}} \put(14,12){\line(1,-2){0.6}}
\multiput(11,16)(1,1){3}{\line(1,-1){1}} \multiput(11,14)(1,1){5}{\line(1,-2){1}} \multiput(12,12)(1,1){6}{\line(1,-1){1}} \multiput(15,13)(1,1){3}{\line(1,-2){1}} \put(17,12){\line(1,-1){1}}

\put(21,11){\circle*{0.5}} \put(22,11){\circle{0.5}} \put(23,11){\circle{0.5}} \put(24,11){\circle*{0.5}} \put(25,11){\circle{0.5}} \put(26,11){\circle{0.5}} \put(27,11){\circle*{0.5}} \put(28,11){\circle{0.5}} \put(21,12){\circle{0.5}} \put(22,12){\circle*{0.5}} \put(23,12){\circle{0.5}} \put(24,12){\circle{0.5}} \put(25,12){\circle*{0.5}} \put(26,12){\circle{0.5}} \put(27,12){\circle{0.5}} \put(28,12){\circle*{0.5}} \put(21,13){\circle{0.5}} \put(22,13){\circle{0.5}} \put(23,13){\circle*{0.5}} \put(24,13){\circle{0.5}} \put(25,13){\circle{0.5}} \put(26,13){\circle*{0.5}} \put(27,13){\circle{0.5}} \put(28,13){\circle{0.5}} \put(21,14){\circle*{0.5}} \put(22,14){\circle{0.5}} \put(23,14){\circle{0.5}} \put(24,14){\circle*{0.5}} \put(25,14){\circle{0.5}} \put(26,14){\circle{0.5}} \put(27,14){\circle*{0.5}} \put(28,14){\circle{0.5}} \put(21,15){\circle{0.5}} \put(22,15){\circle*{0.5}} \put(23,15){\circle{0.5}} \put(24,15){\circle{0.5}} \put(25,15){\circle*{0.5}} \put(26,15){\circle{0.5}} \put(27,15){\circle{0.5}} \put(28,15){\circle*{0.5}} \put(21,16){\circle{0.5}} \put(22,16){\circle{0.5}} \put(23,16){\circle*{0.5}} \put(24,16){\circle{0.5}} \put(25,16){\circle{0.5}} \put(26,16){\circle*{0.5}} \put(27,16){\circle{0.5}} \put(28,16){\circle{0.5}} \put(21,17){\circle*{0.5}} \put(22,17){\circle{0.5}} \put(23,17){\circle{0.5}} \put(24,17){\circle*{0.5}} \put(25,17){\circle{0.5}} \put(26,17){\circle{0.5}} \put(27,17){\circle*{0.5}} \put(28,17){\circle{0.5}} \put(21,18){\circle{0.5}} \put(22,18){\circle*{0.5}} \put(23,18){\circle{0.5}} \put(24,18){\circle{0.5}} \put(25,18){\circle*{0.5}} \put(26,18){\circle{0.5}} \put(27,18){\circle{0.5}} \put(28,18){\circle*{0.5}}

\put(21,17){\line(1,1){1}} \put(21,14){\line(1,1){4}} \put(21,11){\line(1,1){7}} \put(24,11){\line(1,1){4}} \put(27,11){\line(1,1){1}}
\put(21,14){\line(1,-2){1.6}} \put(21,17){\line(1,-2){3}} \put(22,18){\line(1,-2){3.6}} \put(27,11){\line(-1,2){3.6}} \put(28,12){\line(-1,2){3}} \put(28,15){\line(-1,2){1.6}}

\put(31,11){\circle{0.5}} \put(32,11){\circle{0.5}} \put(33,11){\circle*{0.5}} \put(34,11){\circle{0.5}} \put(35,11){\circle{0.5}} \put(36,11){\circle{0.5}} \put(37,11){\circle*{0.5}} \put(38,11){\circle{0.5}} \put(31,12){\circle*{0.5}} \put(32,12){\circle{0.5}} \put(33,12){\circle{0.5}} \put(34,12){\circle{0.5}} \put(35,12){\circle*{0.5}} \put(36,12){\circle{0.5}} \put(37,12){\circle{0.5}} \put(38,12){\circle{0.5}} \put(31,13){\circle{0.5}} \put(32,13){\circle{0.5}} \put(33,13){\circle*{0.5}} \put(34,13){\circle{0.5}} \put(35,13){\circle{0.5}} \put(36,13){\circle{0.5}} \put(37,13){\circle*{0.5}} \put(38,13){\circle{0.5}} \put(31,14){\circle*{0.5}} \put(32,14){\circle{0.5}} \put(33,14){\circle{0.5}} \put(34,14){\circle{0.5}} \put(35,14){\circle*{0.5}} \put(36,14){\circle{0.5}} \put(37,14){\circle{0.5}} \put(38,14){\circle{0.5}} \put(31,15){\circle{0.5}} \put(32,15){\circle{0.5}} \put(33,15){\circle*{0.5}} \put(34,15){\circle{0.5}} \put(35,15){\circle{0.5}} \put(36,15){\circle{0.5}} \put(37,15){\circle*{0.5}} \put(38,15){\circle{0.5}} \put(31,16){\circle*{0.5}} \put(32,16){\circle{0.5}} \put(33,16){\circle{0.5}} \put(34,16){\circle{0.5}} \put(35,16){\circle*{0.5}} \put(36,16){\circle{0.5}} \put(37,16){\circle{0.5}} \put(38,16){\circle{0.5}} \put(31,17){\circle{0.5}} \put(32,17){\circle{0.5}} \put(33,17){\circle*{0.5}} \put(34,17){\circle{0.5}} \put(35,17){\circle{0.5}} \put(36,17){\circle{0.5}} \put(37,17){\circle*{0.5}} \put(38,17){\circle{0.5}} \put(31,18){\circle*{0.5}} \put(32,18){\circle{0.5}} \put(33,18){\circle{0.5}} \put(34,18){\circle{0.5}} \put(35,18){\circle*{0.5}} \put(36,18){\circle{0.5}} \put(37,18){\circle{0.5}} \put(38,18){\circle{0.5}}

\put(31,12){\line(2,-1){2}} \put(31,14){\line(2,-1){6}} \put(31,16){\line(2,-1){7.2}} \put(31,18){\line(2,-1){7.2}} \put(35,18){\line(2,-1){3.2}}
\put(31,16){\line(2,1){4}} \put(31,14){\line(2,1){7.2}} \put(31,12){\line(2,1){7.2}} \put(33,11){\line(2,1){5.2}} \put(37,11){\line(2,1){1.2}}

\put(3.3,9.3){$\rho=\frac12$} \put(13.3,9.3){$\rho=\frac25$} \put(23.3,9.3){$\rho=\frac13$} \put(33.3,9.3){$\rho=\frac14$}

%---------- Second row -------------------------------------------------------

\put(1,1){\circle*{0.5}} \put(2,1){\circle{0.5}} \put(3,1){\circle{0.5}} \put(4,1){\circle{0.5}} \put(5,1){\circle{0.5}} \put(6,1){\circle{0.5}} \put(7,1){\circle*{0.5}} \put(8,1){\circle{0.5}} \put(1,2){\circle{0.5}} \put(2,2){\circle{0.5}} \put(3,2){\circle*{0.5}} \put(4,2){\circle{0.5}} \put(5,2){\circle*{0.5}} \put(6,2){\circle{0.5}} \put(7,2){\circle{0.5}} \put(8,2){\circle{0.5}} \put(1,3){\circle*{0.5}} \put(2,3){\circle{0.5}} \put(3,3){\circle{0.5}} \put(4,3){\circle{0.5}} \put(5,3){\circle{0.5}} \put(6,3){\circle{0.5}} \put(7,3){\circle*{0.5}} \put(8,3){\circle{0.5}} \put(1,4){\circle{0.5}} \put(2,4){\circle{0.5}} \put(3,4){\circle{0.5}} \put(4,4){\circle*{0.5}} \put(5,4){\circle{0.5}} \put(6,4){\circle{0.5}} \put(7,4){\circle{0.5}} \put(8,4){\circle{0.5}} \put(1,5){\circle{0.5}} \put(2,5){\circle*{0.5}} \put(3,5){\circle{0.5}} \put(4,5){\circle{0.5}} \put(5,5){\circle{0.5}} \put(6,5){\circle*{0.5}} \put(7,5){\circle{0.5}} \put(8,5){\circle*{0.5}} \put(1,6){\circle{0.5}} \put(2,6){\circle{0.5}} \put(3,6){\circle{0.5}} \put(4,6){\circle*{0.5}} \put(5,6){\circle{0.5}} \put(6,6){\circle{0.5}} \put(7,6){\circle{0.5}} \put(8,6){\circle{0.5}} \put(1,7){\circle*{0.5}} \put(2,7){\circle{0.5}} \put(3,7){\circle{0.5}} \put(4,7){\circle{0.5}} \put(5,7){\circle{0.5}} \put(6,7){\circle{0.5}} \put(7,7){\circle*{0.5}} \put(8,7){\circle{0.5}} \put(1,8){\circle{0.5}} \put(2,8){\circle{0.5}} \put(3,8){\circle*{0.5}} \put(4,8){\circle{0.5}} \put(5,8){\circle*{0.5}} \put(6,8){\circle{0.5}} \put(7,8){\circle{0.5}} \put(8,8){\circle{0.5}}

\put(1,7){\line(2,1){2}} \put(1,3){\line(1,2){1}} \put(2,5){\line(2,1){2}} \put(4,6){\line(1,2){1}} \put(1,1){\line(2,1){2}} \put(3,2){\line(1,2){1}} \put(4,4){\line(2,1){2}} \put(6,5){\line(1,2){1}} \put(7,7){\line(2,1){1.2}} \put(5,2){\line(-1,-2){0.6}} \put(5,2){\line(2,1){2}} \put(7,3){\line(1,2){1}} \put(7,1){\line(2,1){1.2}}
\put(1,3){\line(2,-1){2}} \put(3,2){\line(1,-2){0.6}} \put(1,7){\line(1,-2){1}} \put(2,5){\line(2,-1){2}} \put(4,4){\line(1,-2){1}} \put(5,2){\line(2,-1){2}} \put(3,8){\line(1,-2){1}} \put(4,6){\line(2,-1){2}} \put(6,5){\line(1,-2){1}} \put(7,3){\line(2,-1){1.2}} \put(5,8){\line(2,-1){2}} \put(7,7){\line(1,-2){1}}

\put(11,1){\circle{0.5}} \put(12,1){\circle*{0.5}} \put(13,1){\circle{0.5}} \put(14,1){\circle{0.5}} \put(15,1){\circle{0.5}} \put(16,1){\circle{0.5}} \put(17,1){\circle*{0.5}} \put(18,1){\circle{0.5}} \put(11,2){\circle{0.5}} \put(12,2){\circle{0.5}} \put(13,2){\circle{0.5}} \put(14,2){\circle{0.5}} \put(15,2){\circle*{0.5}} \put(16,2){\circle{0.5}} \put(17,2){\circle{0.5}} \put(18,2){\circle{0.5}} \put(11,3){\circle{0.5}} \put(12,3){\circle{0.5}} \put(13,3){\circle*{0.5}} \put(14,3){\circle{0.5}} \put(15,3){\circle{0.5}} \put(16,3){\circle{0.5}} \put(17,3){\circle{0.5}} \put(18,3){\circle*{0.5}} \put(11,4){\circle*{0.5}} \put(12,4){\circle{0.5}} \put(13,4){\circle{0.5}} \put(14,4){\circle{0.5}} \put(15,4){\circle{0.5}} \put(16,4){\circle*{0.5}} \put(17,4){\circle{0.5}} \put(18,4){\circle{0.5}} \put(11,5){\circle{0.5}} \put(12,5){\circle{0.5}} \put(13,5){\circle{0.5}} \put(14,5){\circle*{0.5}} \put(15,5){\circle{0.5}} \put(16,5){\circle{0.5}} \put(17,5){\circle{0.5}} \put(18,5){\circle{0.5}} \put(11,6){\circle{0.5}} \put(12,6){\circle*{0.5}} \put(13,6){\circle{0.5}} \put(14,6){\circle{0.5}} \put(15,6){\circle{0.5}} \put(16,6){\circle{0.5}} \put(17,6){\circle*{0.5}} \put(18,6){\circle{0.5}} \put(11,7){\circle{0.5}} \put(12,7){\circle{0.5}} \put(13,7){\circle{0.5}} \put(14,7){\circle{0.5}} \put(15,7){\circle*{0.5}} \put(16,7){\circle{0.5}} \put(17,7){\circle{0.5}} \put(18,7){\circle{0.5}} \put(11,8){\circle{0.5}} \put(12,8){\circle{0.5}} \put(13,8){\circle*{0.5}} \put(14,8){\circle{0.5}} \put(15,8){\circle{0.5}} \put(16,8){\circle{0.5}} \put(17,8){\circle{0.5}} \put(18,8){\circle*{0.5}}

\put(12,1){\line(-2,1){1.2}} \put(11,4){\line(1,2){2}} \put(12,1){\line(1,2){3.6}} \put(18,8){\line(-1,-2){3.6}} \put(17,1){\line(1,2){1}}
\put(11,4){\line(2,-1){6}} \put(18,3){\line(-2,1){7.2}} \put(13,8){\line(2,-1){5.2}}

\put(21,1){\circle{0.5}} \put(22,1){\circle*{0.5}} \put(23,1){\circle{0.5}} \put(24,1){\circle{0.5}} \put(25,1){\circle{0.5}} \put(26,1){\circle{0.5}} \put(27,1){\circle{0.5}} \put(28,1){\circle{0.5}} \put(21,2){\circle{0.5}} \put(22,2){\circle{0.5}} \put(23,2){\circle{0.5}} \put(24,2){\circle*{0.5}} \put(25,2){\circle{0.5}} \put(26,2){\circle{0.5}} \put(27,2){\circle*{0.5}} \put(28,2){\circle{0.5}} \put(21,3){\circle*{0.5}} \put(22,3){\circle{0.5}} \put(23,3){\circle{0.5}} \put(24,3){\circle{0.5}} \put(25,3){\circle{0.5}} \put(26,3){\circle{0.5}} \put(27,3){\circle{0.5}} \put(28,3){\circle{0.5}} \put(21,4){\circle{0.5}} \put(22,4){\circle{0.5}} \put(23,4){\circle*{0.5}} \put(24,4){\circle{0.5}} \put(25,4){\circle{0.5}} \put(26,4){\circle*{0.5}} \put(27,4){\circle{0.5}} \put(28,4){\circle{0.5}} \put(21,5){\circle{0.5}} \put(22,5){\circle{0.5}} \put(23,5){\circle{0.5}} \put(24,5){\circle{0.5}} \put(25,5){\circle{0.5}} \put(26,5){\circle{0.5}} \put(27,5){\circle{0.5}} \put(28,5){\circle*{0.5}} \put(21,6){\circle{0.5}} \put(22,6){\circle*{0.5}} \put(23,6){\circle{0.5}} \put(24,6){\circle{0.5}} \put(25,6){\circle*{0.5}} \put(26,6){\circle{0.5}} \put(27,6){\circle{0.5}} \put(28,6){\circle{0.5}} \put(21,7){\circle{0.5}} \put(22,7){\circle{0.5}} \put(23,7){\circle{0.5}} \put(24,7){\circle{0.5}} \put(25,7){\circle{0.5}} \put(26,7){\circle{0.5}} \put(27,7){\circle*{0.5}} \put(28,7){\circle{0.5}} \put(21,8){\circle*{0.5}} \put(22,8){\circle{0.5}} \put(23,8){\circle{0.5}} \put(24,8){\circle*{0.5}} \put(25,8){\circle{0.5}} \put(26,8){\circle{0.5}} \put(27,8){\circle{0.5}} \put(28,8){\circle{0.5}}

\put(21,3){\line(1,-2){1}} \put(21,8){\line(1,-2){3.6}} \put(24,8){\line(1,-2){3.6}} \put(28,5){\line(-1,2){1.6}}
\put(22,1){\line(2,1){2}} \put(21,3){\line(2,1){2}} \put(26,4){\line(2,1){2}} \put(25,6){\line(2,1){2}} \put(22,6){\line(-2,-1){1.2}} \put(27,2){\line(2,1){1.2}}
\qbezier(24,2)(24.5,2.5)(24.8,2.8) \qbezier(23,4)(23.5,4.5)(23.8,4.8) \qbezier(22,6)(22.5,6.5)(22.8,6.8) \qbezier(27,2)(26.5,1.5)(26.2,1.2) \qbezier(26,4)(25.5,3.5)(25.2,3.2) \qbezier(25,6)(24.5,5.5)(24.2,5.2) \qbezier(24,8)(23.5,7.5)(23.2,7.2) \qbezier(27,7)(27.5,7.5)(27.8,7.8)

\put(31,1){\circle{0.5}} \put(32,1){\circle*{0.5}} \put(33,1){\circle{0.5}} \put(34,1){\circle{0.5}} \put(35,1){\circle*{0.5}} \put(36,1){\circle{0.5}} \put(37,1){\circle{0.5}} \put(38,1){\circle{0.5}} \put(31,2){\circle{0.5}} \put(32,2){\circle{0.5}} \put(33,2){\circle{0.5}} \put(34,2){\circle{0.5}} \put(35,2){\circle{0.5}} \put(36,2){\circle{0.5}} \put(37,2){\circle*{0.5}} \put(38,2){\circle{0.5}} \put(31,3){\circle{0.5}} \put(32,3){\circle{0.5}} \put(33,3){\circle*{0.5}} \put(34,3){\circle{0.5}} \put(35,3){\circle{0.5}} \put(36,3){\circle{0.5}} \put(37,3){\circle{0.5}} \put(38,3){\circle{0.5}} \put(31,4){\circle{0.5}} \put(32,4){\circle{0.5}} \put(33,4){\circle{0.5}} \put(34,4){\circle{0.5}} \put(35,4){\circle*{0.5}} \put(36,4){\circle{0.5}} \put(37,4){\circle{0.5}} \put(38,4){\circle*{0.5}} \put(31,5){\circle*{0.5}} \put(32,5){\circle{0.5}} \put(33,5){\circle{0.5}} \put(34,5){\circle{0.5}} \put(35,5){\circle{0.5}} \put(36,5){\circle{0.5}} \put(37,5){\circle{0.5}} \put(38,5){\circle{0.5}} \put(31,6){\circle{0.5}} \put(32,6){\circle{0.5}} \put(33,6){\circle*{0.5}} \put(34,6){\circle{0.5}} \put(35,6){\circle{0.5}} \put(36,6){\circle*{0.5}} \put(37,6){\circle{0.5}} \put(38,6){\circle{0.5}} \put(31,7){\circle{0.5}} \put(32,7){\circle{0.5}} \put(33,7){\circle{0.5}} \put(34,7){\circle{0.5}} \put(35,7){\circle{0.5}} \put(36,7){\circle{0.5}} \put(37,7){\circle{0.5}} \put(38,7){\circle*{0.5}} \put(31,8){\circle*{0.5}} \put(32,8){\circle{0.5}} \put(33,8){\circle{0.5}} \put(34,8){\circle*{0.5}} \put(35,8){\circle{0.5}} \put(36,8){\circle{0.5}} \put(37,8){\circle{0.5}} \put(38,8){\circle{0.5}}

\put(31,5){\line(2,1){2}} \put(33,6){\line(1,2){1}} \put(32,1){\line(1,2){1}} \put(33,3){\line(2,1){2}} \put(35,4){\line(1,2){1}} \put(36,6){\line(2,1){2}} \put(35,1){\line(2,1){2}} \put(37,2){\line(1,2){1}}
\qbezier(32,1)(31.5,1.5)(31.2,1.8) \qbezier(33,3)(32.5,3.5)(32.2,3.8) \qbezier(35,4)(34.5,4.5)(34.2,4.8) \qbezier(36,6)(35.5,6.5)(35.2,6.8) \qbezier(38,7)(37.5,7.5)(37.2,7.8) \qbezier(33,6)(32.5,6.5)(32.2,6.8) \qbezier(35,1)(34.5,1.5)(34.2,1.8) \qbezier(37,2)(36.5,2.5)(36.2,2.8) \qbezier(38,4)(37.5,4.5)(37.2,4.8)
\qbezier(31,5)(31.5,4.5)(31.8,4.2) \qbezier(31,8)(31.5,7.5)(31.8,7.2) \qbezier(33,6)(33.5,5.5)(33.8,5.2) \qbezier(34,8)(34.5,7.5)(34.8,7.2) \qbezier(33,3)(33.5,2.5)(33.8,2.2) \qbezier(35,4)(35.5,3.5)(35.8,3.2) \qbezier(36,6)(36.5,5.5)(36.8,5.2) \qbezier(37,2)(37.5,1.5)(37.8,1.2)

\put(3.3,-0.7){$\rho=\frac29$} \put(13.3,-0.7){$\rho=\frac15$} \put(23.3,-0.7){$\rho=\frac2{11}$}
\put(33.3,-0.7){$\rho=\frac16$}
\end{picture}
\caption{Ground states configurations for several densities. Occupied sites are denoted by black circles, empty sites by white circles.
Lines are present only to clarify the patterns.}
\label{figconf}
\end{figure}

The situation for densities $\rho\leq\frac12$ that are not mentioned in Theorem \ref{thm2d} is unknown.
All these periodic configurations are present in the grand-canonical phase diagram as
well. Item (b) suggests that the two-dimensional situation is similar to the
one-dimensional one where  a devil's staircase structure may occur. Let us stress that no
periodic configurations occur for large $U$ and densities $\rhoe=\rhoc$ in the intervals
$(\frac16,\frac2{11})$, $(\frac15,\frac29)$, and $(\frac29,\frac14)$. This resembles the
one-dimensional situation, but for small $U$.

\bfig
\epsfxsize=100mm
\centerline{\epsffile{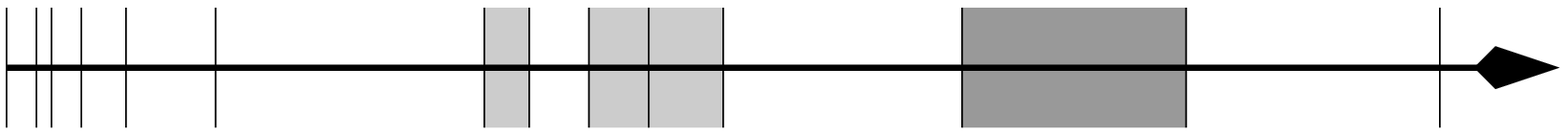}}
\figtext{
\writefig	12.5	1.1	{$\rhoe$}
\writefig	2.3	0.35	{0}
\writefig	2.97	0.35	{$\frac1{25}$}
\writefig	3.55	0.35	{$\frac1{13}$}
\writefig	5.32	0.35	{$\frac16$}
\writefig	5.55	0.35	{$\frac2{11}$}
\writefig	6.03	0.35	{$\frac15$}
\writefig	6.43	0.35	{$\frac29$}
\writefig	6.9	0.35	{$\frac14$}
\writefig	8.4	0.35	{$\frac13$}
\writefig	9.85	0.35	{$\frac25$}
\writefig	11.45	0.35	{$\frac12$}
}
\caption{Canonical ground state phase diagram in two dimensions for $U>8$.}
\label{figphdcan2d}
\end{figure}

\section*{Bibliographical notes}

The Falicov-Kimball model was introduced in \cite{FK}, and the first rigorous results
appeared in \cite{BS} and \cite{KL}.
A simple derivation of the expansion \eqref{expgsen} using Cauchy formula can be found in \cite{GM}. It can be extended to positive temperatures with the help of
Lie-Schwinger series \cite{DFF}. Segregation in arbitrary dimension (Theorem \ref{thmseg}) was proposed in
\cite{FLU}. Domains in the grand-canonical
phase diagram for the empty, full, and chessboard configurations, were obtained using lower
bounds involving correlation functions of classical particles, see \cite{GM} for
references. Eq.\ \eqref{esmallU} for the ground state
energy for $U$ small was derived using Wigner-Brillouin degenerate perturbation theory in
\cite{FGM}. For the results in two dimensions (Theorem \ref{thm2d}), we refer to \cite{HK} and references
therein. Results on the ground state for $\mue$ in the universal gap have been extended to positive
temperatures \cite{DFF} using ``quantum Pirogov-Sinai theory'' (see the review in this
Encyclopedia). Further studies and extensions of the
Falicov-Kimball model include interfaces, non-bipartite lattices, bosonic
particles, continuous fields instead of classical particles, magnetic fields, correlated
and extended hoppings.
An extensive list of references can be found in the reviews \cite{FZ,GM,JL}.

\bigskip
{\bf See also:} Equilibrium Statistical Mechanics. Quantum Statistical Mechanics. Fermionic Systems.
Hubbard Model. Pirogov-Sinai Theory.

\end{document}